\documentclass[10pt]{article}
\usepackage[letterpaper]{geometry}
\usepackage{hyperref}
\providecommand{\keywords}[1]{\textbf{\textit{Keywords:}} #1}

\usepackage{graphicx}
\usepackage{pdfpages}			
\usepackage{lipsum}				
\usepackage{amsfonts,amsmath,amssymb,latexsym}
\usepackage{mathrsfs}
\usepackage{enumerate}
\usepackage{enumitem}
\usepackage{mathtools}
\usepackage[normalem]{ulem}

\usepackage{algorithm}
\usepackage{algpseudocode} 
\usepackage{algorithmicx}


\algnewcommand{\Input}[1]{\Statex \textbf{input: } #1 }
\algnewcommand{\Output}[1]{\Statex \textbf{output: } #1 \Statex }
\algnewcommand{\return}[1]{\State\textbf{return} #1}
\makeatletter
\newenvironment{breakablealgorithm}
{
	\begin{center}
		\refstepcounter{algorithm}
		\hrule height.8pt depth0pt \kern2pt
		\renewcommand{\caption}[2][\relax]{
			{\raggedright\textbf{Algoritmo~\thealgorithm} ##2\par}%
			\ifx\relax##1\relax 
			\addcontentsline{loa}{algorithm}{\protect\numberline{\thealgorithm}##2}%
			\else 
			\addcontentsline{loa}{algorithm}{\protect\numberline{\thealgorithm}##1}%
			\fi
			\kern2pt\hrule\kern2pt
		}
	}{
		\kern2pt\hrule\relax
	\end{center}
}
\makeatother

\usepackage{tikz}
\usetikzlibrary{shapes,shapes.multipart, calc,matrix,arrows,positioning,automata,snakes}
\usepackage{subfig}

\newcommand{\yI}{\mathcal{I}}  
\newcommand{\yS}{\mathcal{S}}  
\newcommand{\yA}{\mathcal{A}}  
\newcommand{\yQ}{\mathcal{Q}}  
\newcommand{\yR}{\mathcal{R}} 
\newcommand{\yB}{\mathcal{B}}  
\newcommand{\yC}{\mathcal{C}}

\newtheorem{theorem}{Theorem}

\tikzset{
	>=stealth',
	punkt/.style={
		circle,
		rounded corners,
		draw=black, thick,
		text width=1.5em,
		minimum height=3em,
		text centered},
	punkts/.style={
		circle,
		rounded corners,
		draw=black, thick, 
		text width=1em,
		minimum height=1em,
		text centered},
	invisible/.style={
		draw=none,
		text width=1.5em,
		minimum height=0em,
		text centered},
	inv/.style={
		draw=none,
		text width=2.5em,
		minimum height=3em,
		text centered},
	pil/.style={
		->,
		thick,
		shorten <=2pt,
		shorten >=2pt,}
}

\newcommand{\ioco}{\text{\bf ioco}\,\,}
\newtheorem{definition}{Definition}
\newtheorem{prop}{Proposition}
\newtheorem{lemma}{Lemma}

\begin{document}

\title{Automatically Checking Conformance on Asynchronous Reactive Systems}

\author{Camila Sonada Gomes\thanks{Computing Department, University of Londrina, Londrina, Brazil.} \and 
	Adilson Luiz Bonifacio\thanks{Computing Department, University of Londrina, Londrina, Brazil.}}

\date{} 

\maketitle

\begin{abstract}
	Software testing is one of the most important issues in the software development process to ensure the quality of the products. 
	Formal methods have been promising on testing reactive systems, specially critical systems, where accuracy is mandatory  and any fault can cause severe damage. 
	Systems of this nature are characterized by receiving messages from the environment and producing outputs in response.
	One of the biggest challenges in  model-based testing is the  conformance checking of  asynchronous reactive systems modeled by IOLTSs. 
	The aim is to verify if an implementation is in compliance with its respective specification. 
	In this work, we develop a practical tool to check the conformance relation  between reactive models using a more general theory based on regular languages. 
	Our tool also offers the \ioco conformance checking and an intuitive interface to apply practical experiments. 
	In addition, we present 
	some testing scenarios in practical applications and compare to other tools from the literature using both  notions  of conformance. 
	
\end{abstract}

	\section{Introduction}
	
	Automatic testing tools have been proposed to support the  development process of reactive systems that are characterized by  continuous interaction with the environment.
	In this setting, systems receive external stimuli and produce outputs, asynchronously, in response. 
	In addition, systems of this nature are usually critical and require more accuracy in their development process, especially in the testing activity, where appropriate formalisms must be used as the basis~\cite{bonifacio2018,simao2014,tretmans2008}.
	Input Output Labeled Transition Systems (IOLTSs)~\cite{simao2014,tretmans2008,aichernig2016,tretmans1999} are 
	traditional formalisms usually applied to model and test reactive systems.

	In model-based testing, an IOLTS specification can model desirable and undesirable behaviors of an implementation under test (IUT). 
	The aim is to find faults in an IUT according to a certain fault model~\cite{tretmans1996,bonifacio2018,aichernig2008} in order to show if requirements are satisfied regarding its respective system specification. 
	A well-established conformance relation, called \ioco and proposed by Tretmans~\cite{tretmans1996}, requires that the outputs produced by an IUT must also be produced by its respective specification, following  the same behavior. 
	A more recent and general conformance relation, proposed by Bonifacio and Moura~\cite{bonifacio2018}, specifies desirable and  undesirable behaviors using regular languages to define the testing fault model.
	
	In this work, we address the development of an automatic tool for  conformance verification of asynchronous reactive systems modeled by IOLTSs. 
	Here we introduce in a practical tool both notions of conformance in order to  provide a wider application range compare to other tools. 
	JTorx~\cite{jtorx}, for instance, is a tool from the literature that also implements a conformance testing verification process, but only based on the classical \ioco relation. 
	Our tool comprises both the classical \ioco relation and also the more general conformance based on regular languages. 
	We also run some practical sceneries 
	to evaluate aspects related to the effectiveness and usability of both conformance theories and these tools.

	We organize this paper as follows.
	Section~\ref{sub:metodo} describes the conformance verification methods using regular languages and the  \ioco relation. 
	The practical tool which implements the more general method of conformance checking is presented in Section~\ref{sub:ferramenta}. 
	Section~\ref{sub:comparative} describes the comparative analysis of tools.
	Some applications and a comparative study are given in Section~\ref{sub:aplicacao}.  
	Section~\ref{sub:conclusao} offers some concluding remarks.

	\section{Conformance Verification}\label{sub:metodo}
	
	The conformance checking task can determine if an IUT is in compliance with its specification when both are modeled by appropriate formalisms. 
	The classical \ioco conformance relation~\cite{tretmans1999, tretmans2008} establishes the compliance between IUTs and specifications when they are specified by IOLTS~\cite{tretmans1999, simao2014, tretmans2008,aichernig2016}.
	An IOLTS  is a variation of the Labeled Transition Systems (LTS)~\cite{tretmans1992, daca2014, zeng2009, cartaxo2007} with the partitioning of input and output labels. 
	\begin{definition}
		\label{def:iolts}
		An IOLTS is given by $\yS = (S, s_0, L_I, L_U, T)$ where: 
		\begin{itemize}
			\item $S$ is the set of states;
			\item $s_0 \in S$ is the initial state;
			\item $L_I$ is a set of input labels;
			\item $L_U$ is a set of output labels;
			\item $L=L_I \cup L_U$ and $L_I \cap L_U =\emptyset$;
			\item $T$ $\subseteq S \times (L \cup \{ \tau \}) \times S$ is a finite set of transitions, where the internal action $\tau \notin L$; and 
			\item $(S, s_0, L, T)$ is the underlying LTS associated with $\yS$.	
		\end{itemize}	
	\end{definition}
	A transition $(s, l, s')\in T$ indicates that from the state $s \in S$ with the label $l \in (L \cup \{ \tau \})$ the state $s' \in S$ is reached in an LTS/IOLTS model.
	When we have a transition $(s, \tau, s')\in T$ with an internal action it means that 
	an external observer can not see the movement from state $s$ to state $s'$ in the model. 
	
	
	We may also have the notion of quiescent states in IOLTS models. 
	If a state $s$ of an IOLTS has no output $x\in L_U$ and internal action $\tau$ defined on it, we say that $s$ is quiescent~\cite{tretmans2008}. 
	When a state $s$ is quiescent we then add a transition $(s,\delta,s)$, where 
	$\delta \notin L_ \tau$\footnote{In order to ease the notation we denote $L \cup \{\tau \}$ by $L_\tau$.}. 
	In a real scenario of black-box testing where an IUT sends messages to a tester and receives back responses, quiescence indicates that an IUT could no longer respond to the tester, or it has timed out, or even it is simply slow. 
	
	We will need the semantics of LTS/IOLTS models, but first we introduce the notion of paths. 
	\begin{definition}(\cite{bonifacio2018})
		Let $\yS = (S, s_0, L, T)$ be a LTS and $p,q \in S$. 
		Let $\sigma = l_1, \cdots, l_n$ be a word in $L^\star_\tau$. 
		We say that $\sigma$ is a \emph{path} from $p$ to $q$ in $\yS$ if there are
		states $r_i \in S$, 
		and labels $l_i \in L_\tau$, $1 \leq i \leq n$, such that $(r_{i-1}, l_i, r_i) \in T$, $1 \leq i \leq n$, with $r_0=p$ and $r_n = q$.		 
		We also say that $\alpha$ is an \emph{observable path} from $p$ to $q$ 
		in $\yS$ if we remove  the internal actions $\tau$ from $\sigma$. 
	\end{definition}
	A path can also be 
	denoted by $s \xrightarrow[]{\sigma} s'$, where the behavior $\sigma \in L^\star_\tau$  starts in the state $s \in S$ and reaches the state $s' \in S$. 
	An observable path $\sigma$, from $s$ to $s'$, is denoted by $s\xRightarrow[]{\sigma} s'$. 
	We can also write $s \xrightarrow[]{\sigma} $ or $s\xRightarrow[]{\sigma} $ when the final state is not important. 
	
	Paths that start from state $s$ are called paths of $s$, and the semantics of an LTS model is given by the paths that start from their initial state.
	The semantics of the LTS model are given below.
	\begin{definition}\label{def:traceSemantica}(\cite{bonifacio2018}).
		Let $\yS = (S, s_0, L, T)$ be a LTS  and $s \in S$:
		\begin{enumerate}
			\item The set of paths of $s$ is given by $tr(s) =  \{\sigma \arrowvert s \xrightarrow[]{\sigma} \} $ and the set of observable paths of $s$ is 
			$otr(s) = \{ \sigma \arrowvert s \xRightarrow[]{\sigma} \} $. 
			\item The semantics of $\yS$ is $tr(s_0)$ or $tr(\yS)$ and the observable semantics of $\yS$ is $otr(s_0)$ or $otr(\yS)$.
		\end{enumerate}
	\end{definition}
	The semantics of an IOLTS is defined by the semantics of the underlying LTS.

	
	Conformance checking can be established between IOLTS models over the \ioco relation. 
	When we apply input stimuli
	to both a specification and an IUT, if the IUT produces outputs that are also defined in the specification, 
	we say that the IUT conforms to the specification. 
	Otherwise, we say that they do not conform~\cite{tretmans2008}.
	\begin{definition}(\cite{tretmans2008,bonifacio2018}).
		Let $\yS = (S, s_0, L_I, L_U, T)$ be a specification  and  $\yI = (Q, q_0, L_I, L_U, R)$ be an IUT. 
		We say that  
		$ \yI \; ioco \; \yS$  if, and only if,  $out(q_0 \; after \; \sigma) \subseteq out(s_0 \; after \; \sigma)$ for all $ \sigma \in otr(\yS)$, where $s \; after \; \sigma = \{q| s \xRightarrow[]{\sigma} q\}$ for all $s \in S$ and all $\sigma \in otr(\yS)$, and the function $out(V) = \bigcup\limits_{s\in V}^{} \{l \in L_U \arrowvert s \xRightarrow[]{l}\}$.	
	\end{definition}
	
	On the other hand, a 
	more general conformance relation has been proposed by Bonifacio and  Moura~\cite{bonifacio2018}, where the fault model is defined by regular languages. This approach provides a wider fault coverage for both LTS and IOLTS models.
	Basically, desirable and undesirable behaviors are specified by regular expressions, $D$ and $F$, respectively. 
	Given an implementation $\yI$, a specification $\yS$, and regular languages $D$ and  $F$, 
	$\yI$ is in compliance with 
	$\yS$ according $(D, F)$, i.e, $\yI conf_{D,F} \;\yS$ if, and only if, no undesirable behavior of $F$ is observed in $\yI$ and is specified in $\yS$, and all desirable behaviors of $D$ are observed in $\yI$ and also are specified in $\yS$.
	\begin{definition}\label{definicao:confLing}(\cite{bonifacio2018})
		Let a set of symbols $L$, and the languages $\mathcal{D,F} \subseteq L^\star$ over $L$. 
		Let $\yS$ and $\yI$, LTS models, with $L$ as their set of labels, $\yI conf_{D,F} \;\yS $ if, and only if: 
		\begin{enumerate}
			\item $ \sigma \in otr(\yI) \cap F$,  then  $\sigma \notin otr(\yS)$; and 
			\item $ \sigma \in otr(\yI) \cap D$, then  $\sigma \in otr(\yS)$.
		\end{enumerate}
	\end{definition}
	
	We remark that ordinary LTS models can be checked using the language-based conformance approach using only the notion of desirable and undesirable behaviors. 
	In this case, we do not need to partition the alphabet into input and output labels, as required by IOLTS models and crucial for  \ioco relation.

	\begin{prop}\label{prop:verifConf}(\cite{bonifacio2018}).
		Let the specification $\yS$ and the IUT $\yI$ be  LTS models over $L$, and the languages $D,F \subseteq L^\star$ over $L$. 
		We say that $\yI \;conf_{D,F}\; \yS$ if, and only if, 
		$otr(\yI) \cap [(D \cap \overline{otr}(\yS)) \cap (F \cap otr(\yS))] = \emptyset$, where $\overline{otr}(\yS)$ is the complement of $otr(\yS)$ given
		by $\overline{otr}(\yS) = L^\star - otr(\yS)$.
	\end{prop}
	
	Lemma~\ref{lema:iocoLing} shows that 
	the more general notion of conformance relation given in Definition~\ref{definicao:confLing} restrains the classical \ioco conformance relation. 
	\begin{lemma}\label{lema:iocoLing}(\cite{bonifacio2018}).
		Let a specification $\yS = (S, s_0, L_I, L_U, T)$ and an IUT $\yI = (Q, q_0, L_I, L_U, R)$ be IOLTS models, we have that 
		$\yI \; ioco \; \yS$ if, and only if, $\yI \; conf_{D,F} \; \yS$ when $D=otr(\yS)L_U$ e $F=\emptyset$. 
	\end{lemma}
	Clearly, the \ioco relation can be given by the more general conformance relation using regular languages. 

	
	In what follows we will see that the conformance checking can be obtained using the automata theory. 
	But first, LTS/IOLTS models must be transformed into Finite State Automatons~(FSAs) as proposed by Bonifacio and Moura~\cite{bonifacio2018}. 
	Hence, the union, intersection, and complement operations over regular languages can be used in the conformance verification process. 
	
	We start by formally defining FSA models~\cite{sipser2006}. 
	\begin{definition}\label{def:fsa}
		A FSA $\mathcal{M}$ is defined as $(S, s_0, L, T, F)$ where:
		\begin{itemize}
			\item $S$ is the set of states;
			\item $s_0$ is the initial state;
			\item $L$ is the alphabet;
			\item $T \subseteq S \times (L \cup \{\epsilon\}) \times S$ is the finite set of transitions,  $\epsilon \notin L$ represents empty \textit{string};
			\item $F \subseteq S$ is the set of final states or acceptance states.
		\end{itemize}
	\end{definition}
	
	Any LTS/IOLTS $\yS$ can be transformed into a FSA  $\mathcal{M}$
	by mapping 
	internal actions $\tau$ of $\yS$ into $\epsilon$-transitions in  $\mathcal{M}$, and all states of $\yS$ are defined as final states in  $\mathcal{M}$.
	\begin{definition}(\cite{bonifacio2018})
		Let $\yS = (S, s_0, L, T)$ be a LTS. The FSA induced by $\yS$ is $\yA_{\yS} = (S, s_0,L, \rho, S)$ where, for all $p,q \in S$ and all $l \in L$, we have:
		\begin{itemize}
			\item $(p, l, q) \in \rho$ if and only if $(p, l, q) \in T$;		
			\item $(p, \epsilon, q) \in \rho$ if and only if $(p, \tau, q) \in T$. 
		\end{itemize}
	\end{definition}
	
	The semantics of a FSA is given by the language it accepts. 
	A language $R\subseteq L^\star$ is \emph{regular} if there is a FSA $\mathcal{M}$  such that $L(\mathcal{M})=R$ where $L$ is an alphabet~\cite{sipser2006}.
	Since $D$ and $F$ are regular languages, we can effectively construct the 
	automatons  $\yA_{D}$  and $\yA_{F}$  such that $D = L(\yA_D)$ and $F = L(\yA_F)$. 
	
	%

	The following proposition says how to obtain FSAs that capture the union, the intersection and the complement of regular languages  accepted by their respective automatons.
	\begin{prop}(\cite{bonifacio2018})\label{prop:operations}
		Let $L$ be a set of symbols, and let $\yA = (S_\yA, a_0, L, \rho_\yA, F_\yA)$ and $\yB = (S_\yB, b_0, L, \rho_\yB, F_\yB)$ be FSAs. Then we can effectively construct a FSA $\mathcal{T} = (S_\mathcal{T}, t_0, L, \rho_\mathcal{T}, F_\mathcal{T})$ such that 
		\begin{enumerate}
			\item $L(\mathcal{T}) = L(\yA) \cap L(\yB)$;
			\item $L(\mathcal{T}) = L(\yA) \cup L(\yB)$;
			\item $L(\mathcal{T}) = \overline{L(\yA)} = L^* - L(\yA)$. 
		\end{enumerate}
	\end{prop}
	
	
	
	Next, we introduce the notions of the test case and test suite according to formal languages. 
	\begin{definition}(\cite{bonifacio2018}).
		Let a set of symbols $L$, the test suite $T$ over $L$ is a language, where $T \subseteq L^\star$, so that each $\sigma \in T$ is a test case.
	\end{definition}
	If the test suite is a regular language, then there is a  FSA $\mathcal{A}$ that accepts it, such that the final states of  $\mathcal{A}$ are fault states.
	The set of undesirable behaviors, defined by these fault  states, is called by fault model of $\yS$~\cite{bonifacio2018}.
	
	Lemma~\ref{lema:conjTestComp} says that a 
	complete test suite can be obtained from an IOLTS specification $\yS$ and a pair of languages $(D,F)$ using the Proposition~\ref{prop:verifConf}. 
	The test suite is able to identify the absence of desirable behaviors specified by $D$ and the presence of undesirable behaviors specified by $F$ in the specification $\yS$. 
	\begin{lemma}\label{lema:conjTestComp}(\cite{bonifacio2018}).
		Let an alphabet $L$, a specification LTS $\yS$ and regular languages $D,F \subseteq L^\star$. 
		The only complete test suite 
		for $\yS$ using $(D,F)$ is given by:
		\[
		T = [(D\cap \overline{otr}(\yS)) \cup (F \cap otr(\yS))].
		\]
	\end{lemma}
	
	
	We declare that an implementation $\yI$ is in compliance with a specification $\yS$   
	if there is no test case of the test suite $T$ that is also a behavior of $\yI$~\cite{bonifacio2018}. 
	\begin{definition}(\cite{bonifacio2018}).
		Let the test suite $T$ and a IUT $\yI$. 
		$\yI$ adheres to $T$ if, and only if, for all $\sigma \in otr(\yI)$ have to $\sigma \notin T$. 
	\end{definition}

	We also need the notion of determinism over LTS/IOLTS models. 
	\begin{definition}
		Let $\yS = (S, s_0, L, T)$ be a LTS. We say that $\yS$ is deterministic if  $\yS$ has no $\tau$-labeled transitions and, $s_0 \xRightarrow[]{\sigma} s_1$ and $s_0 \xRightarrow[]{\sigma} s_2$ imply $s_1 = s_2$ for all $s_1,s_2 \in S$ and all $\sigma \in L^*$.
	\end{definition}
	
	Let $\yS$ be a deterministic IOLTS, 
	so the automaton $\yA_1$ induced by $\yS$ is 
	also deterministic,  
	and we write $L(\yA_1) = otr(\yS)$. 
	Hence, we can effectively obtain a 
	FSA $\yA_2$ 
	such that $L(\yA_2) = L(\yA_F) \cap L(\yA_1) = F \cap otr(S)$ by Proposition~\ref{prop:operations}.

	Consider the FSA $\yB_1$ obtained from $\yA_1$ by reversing its set of final states, that is, a state $s$ is a final state in $\yB_1$ if and only if $s$ is not a final state in $\yA_1$. Clearly, $L(\yB_1) = \overline{L(\yA_1)} = \overline{otr}(\yS)$. 
	We can now effectively get a FSA $\yB_2$ 
	such that $L(\yB_2) = L(\yA_D) \cap L(\yB_1) = D \cap \overline{otr}(\yS)$.
	
	Since $\yA_2$ and $\yB_2$ are FSAs, 
	we can construct 
	a FSA $\yC$ 
	such that $L(\yC) = L(\yA_2) \cup L(\yB_2)$, where 
	$L(\yC) = T$.
	We can conclude that when $D$ and $F$ are regular languages and $\yS$ is a deterministic specification,
	then a complete FSA $\mathcal{T}$ can be constructed 
	such that $L(\mathcal{T}) = T$. 
	
	\begin{prop}(\cite{bonifacio2018})
		Let $\yS$ and $\yI$ be the deterministic specification and implementation IOLTSs over $L$
		with $n_S$ and $n_I$ states, respectively. Let also $|L| = n_L$. Let $\yA_D$ and $\yA_F$ be deterministic FSAs
		over $L$ with $n_D$ and $n_F$ states, respectively, and such that $L(\yA_D) = D$ and $L(\yA_F) = F$. Then,
		we can effectively construct a complete FSA $\mathcal{T}$ with $(n_S + 1)^2n_Dn_F$ states, and such that $L(\mathcal{T})$
		is a complete test suite for $\yS$ and $(D, F)$. Moreover, there is an algorithm, with polynomial time
		complexity $\Theta(n^2_Sn_In_Dn_Fn_L)$ that effectively checks whether $\yI conf_{D,F} \yS$ holds.
	\end{prop}
	
	We obtain a similar result for the \ioco conformance relation using Lemma~\ref{lema:iocoLing}. 
	\begin{theorem}(\cite{bonifacio2018})
		Let $\yS$ and $\yI$ be deterministic specification and implementation IOLTSs over $L$ with
		$n_S$ and $n_I$ states, respectively. Let $L = L_I \cup L_U$, and $|L| = n_L$. Then, we can effectively construct
		an algorithm with polynomial time complexity $\Theta(n_Sn_In_L)$ that checks whether $\yI$ \ioco $\yS$ holds.
	\end{theorem}

	\section{A Testing Tool for Reactive Systems}\label{sub:ferramenta} 
	
	In this work, we have developed 
	the automatic checking conformance tool \emph{Everest}\footnote{EVEREST -- \textit{conformancE Verification on tEsting ReactivE SysTems} 
		\\ Available in \url{https://everest-tool.github.io/everest-site}} 
	for reactive systems modeled by LTS/IOLTS. 
	Our tool supports the more general notion of conformance based on regular languages and also the classical notion of \ioco relation. 
	\emph{Everest} has been developed in Java~\cite{java} using 
	the \textit{Swing} library~\cite{javaSwing}, providing a yielding and friendly usability experience through a graphical interface. 
	
	Some features provided by the Everest tool are: 
	(i)  check conformance  based on regular languages and \ioco relation;
	(ii)  describe desirable and undesirable behaviors using regular expressions;
	(iii) specify formal  models in Aldebaran format~\cite{aut}; 
	(iv) generate test suites when non-conformance verdicts are obtained;
	(v)  provide  state paths, i.e, the sequence of states induced by a test case over the IUT and specification; and 
	(vi) allow the graphical representation of the formal  models.
	
\begin{figure}[hbt]
	\centering
	\includegraphics[scale=0.35]{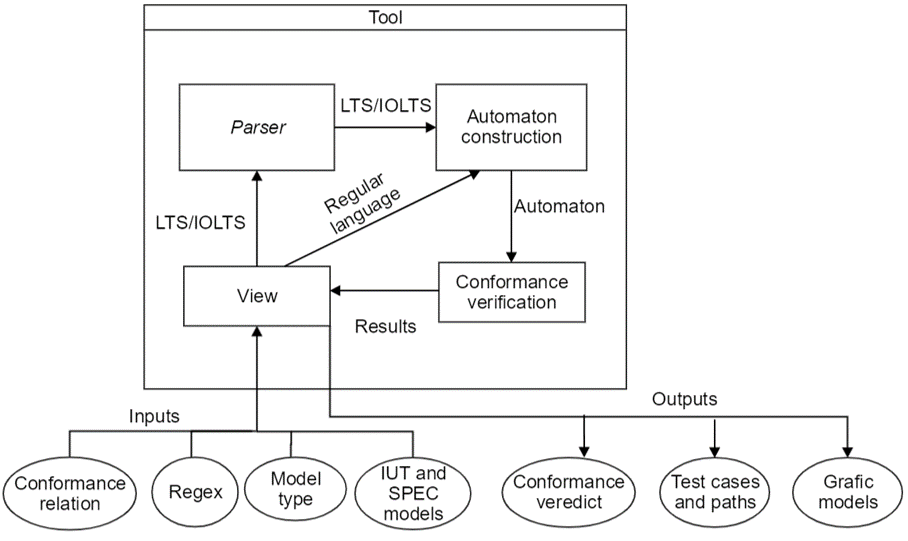}
	\caption{Tool's Architecture}
	\label{fig:arquiteturaFerramenta}
\end{figure}

	The tool's architecture is organized in four modules as depicted in 
	Figure~\ref{fig:arquiteturaFerramenta}. 
	The modules are given by rectangles and  the data flow between them is denoted by the arrows. 
	The input data and the output results are represented by ellipses. 
	
	The \textit{View} module implements an intuitive graphical interface with three different views:  configuration; \ioco conformance; and language-based conformance. 
	In the configuration view we set the specification and implementation models (in Aldebaran format~\cite{.aut}), the model type (LTS or IOLTS) under test, and the partition into input and output labels for IOLTS models, if it is the case. 
	In the \ioco and language-based conformance views, we run both  verification processes and also graphically inspect the models to ascertain the configuration information.
	After finishing the testing process, if the IUT does not conform to the specification then a negative verdict is displayed together with the associated test cases. 
	Otherwise, the tool displays a positive verdict of conformance. 
	Note that we can provide regular expressions to represent the  desirable behaviors and the fault properties in the language-based conformance view. 
	
	The IUT and specification models are validated by the \textit{Parser} module where data structures are constructed to internally represent them.
	The \textit{Automaton Construction} module transforms the LTS/IOLTS models into their respective finite automatons which, in turn, are used to construct the fault model together with the automatons obtained by means of regular languages. 
	
	
	The \textit{Conformance Verification} module  provides all necessary operations  over regular languages such as union, intersection, and complement~\cite{sipser2006}. 
	This module also constructs the finite automaton that represents the complete test suite and 
	comprises both conformance verification techniques. 
	
	Next, we describe the conformance checking processes and their essential algorithms. 
	Some other basic algorithms are not described here since they are given by classical constructions on the automata theory, 
	such as intersection, complement, and union operations over regular languages. 
	In the same vein, we omit the conversion algorithm from regular languages into the automatons that accept them, the conversion algorithm from LTS/IOLTS models into the associated FSAs, and the automaton determinization algorithm. 
	
	
	\subsection{\ioco Conformance Verification}
	
	The Algorithm~\ref{algoritmo:ConfIOCO} performs the \ioco verification given the IOLTS models of the specification $\yS$ and the IUT $\yI$. 
	The line $2$ of the algorithm indicates the fault model construction for the specification $\yS$ represented by the automaton $\yA_{T}$. 
	At line $3$ the IUT $\yI$ is transformed into the automaton  $\yA_{I}$.
	Lines $4$ and $5$ build the automaton $\yA_{B}$ which recognizes the language obtained from the intersection of the languages accepted by the automatons $\yA_{T}$ and $\yA_{I}$. 
	Note that $\yA_{T}$ represents the fault model that captures the behaviors that should not be found in the IUT.
	When this intersection is not empty it means that some behavior is present in both the fault model and the IUT.
	In this case, the behavior is a word accepted by the automaton $\yA_{B}$  thus its 
	set of acceptance states $E$ it not empty.
	Then, from line $6$ to line $10$ whenever the set $E$ of final states is empty the IUT $\yI$ is declared in compliance to the specification $\yS$. 
	Otherwise, a fault is detected and $\yI$ is declared non-conforming to the specification $\yS$. 
	\hfill\\
	\begin{breakablealgorithm}
		\label{algoritmo:ConfIOCO}
		\caption{Verify \ioco Conformance}	
		\begin{algorithmic}[1]
			\Input{IOLTSs $\yS, \yI$} 
			\Output{$\yI$ \ioco conforms to $\yS$, or not}
			\Function{\textsc{verifyIOCOConformance$(\yS, \yI)$}}{}
			\State $\yA_{T} = faultModelIoco(\yS)$ 
			\State $\yA_{I} = ioltsToAutomaton(\yI)$ 
			\State $\yA_{B} =  (S_B,s_0, L, \delta, E)$
			\State $\yA_{B} = intersection(\yA_{T}, \yA_{I})$ 
			
			\If{$E=\emptyset$}
			\return{$\yI$ \ioco $\yS$}
			\Else
			\return{$\yI$ \sout{ioco} $\yS$}
			\EndIf				
			\EndFunction				
		\end{algorithmic}
	\end{breakablealgorithm}
	\hfill\\
	The Algorithm~\ref{algoritmo:TS_IOCO}  constructs  the fault model automaton $\yA_{T}$ based on specification $\yS$. The fault model is the set of behaviors that should not be found in the implementation.
	The construction of this model is based on Lemma~\ref{lema:conjTestComp}, where the sets of behaviors $D$ and $F$ are defined as $D = otr(\yS)L_U$ and $F = \emptyset$. 
	
	At line $2$ of the algorithm is constructed the automaton that recognizes the behaviors in the specification that ends with the production of an output. 
	At line $3$ the automaton $\yA_{S}$ is obtained induced by the specification $\yS$. 
	At line $4$ the automaton $\yA_{compS}$ is constructed to accept the behaviors that are not present in the language accepted by the specification automaton.
	The fault model is then obtained at line $5$, where automaton $\yA_{T}$ is constructed in order to capture the common behaviors of the languages accepted by the automaton $\yA_{D}$ e $\yA_{compS}$. 
	Thus, the fault model represents the behaviors considered as faults in candidate implementations.
	\hfill\\ 
	\begin{breakablealgorithm}
		\label{algoritmo:TS_IOCO}
		\caption{Fault Model \ioco}
		
		\begin{algorithmic}[1]		
			\Input{IOLTS $\yS$} 
			\Output{The fault model automaton}	
			\Function{\textsc{faultModelIoco$(\yS)$}}{}	
			\State $\yA_{D} = modelD(\yS)$ 	
			\State $\yA_{S} = ioltsToAutomaton(\yS)$ 	
			\State $\yA_{compS} = complement(\yA_{S})$ 			
			\State $\yA_{T} = intersection(\yA_{D}, \yA_{compS})$ 
			\return{$\yA_{T}$}	
			\EndFunction	
		\end{algorithmic}
	\end{breakablealgorithm}
	\hfill\\
	The construction of an automaton that accepts the specification behaviors that end with an output is obtained by Algorithm~\ref{algoritmo:ConstroiD}. 
	The automaton $\yA_{D}$ is constructed to capture
	the behaviors defined by the set $D = otr(\yQ)L_U$. 
	
	The states set of $\yA_{D}$ is  defined at  line $3$, by the states set of $\yQ$ and the special state $d$. 
	The state $d$ then must be the only acceptance state (line $6$) of $\yA_{D}$. 
	The transitions of $\delta$ are created as follows:
	\begin{enumerate}
		\item Lines $7$ to $13$ transform transitions $\tau$ of IOLTS $\yQ$ in transitions $\epsilon$ of automaton $\yA_{D}$; 
		or in transitions with the same label. 
		\item Lines $14$ to $18$ create new transitions in $\delta$ with output labels $L_U$  from every state of $q \in Q$ to the special state $d$. 
	\end{enumerate}
	The resulting automaton $\yA_{D}$ can be non-deterministic, then line $19$ returns a deterministic automaton using the function \textit{convertToDeterministicAutomaton}.
	\hfill\\
	\begin{breakablealgorithm}
		\label{algoritmo:ConstroiD}
		\caption{Model D} 
		\begin{algorithmic}[1]	
			\Input{IOLTS $\yQ = (Q, q_0, L_I, L_U, T)$}
			\Output{$\yA_{D}$ automaton}
			
			\Function{\textsc{modelD$(\yQ)$}}{}
			\State $\yA_{D} = (Q,q_0, L, \delta, F)$
			\State $Q = Q \cup d$
			\State $L = L_I \cup L_U$
			\State $\delta = \emptyset$
			\State $F = d$
			\For{$(p,x,r) \in T$}
			\If{$x = \tau$}
			\State $\delta = \delta \cup (p,\epsilon,r)$
			\Else
			\State $\delta = \delta \cup (p,x,r)$			
			\EndIf
			\EndFor
			
			\For{$q \in Q$}
			\For{$l \in L_U$}
			\State $\delta = \delta \cup (q,l,d)$
			\EndFor	
			\EndFor		
			\return{$convertToDeterministicAutomaton(\yA_{D})$}
			\EndFunction						
		\end{algorithmic}
	\end{breakablealgorithm}
	
	
	\subsection{Based-Language Conformance Verification}
	
	The language-based conformance verification is performed by Algorithm~\ref{algoritmo:ConfLing}. 
	Let an IUT $\yI$, a specification $\yS$ and the pair of languages $(D,F)$, the algorithm checks whether $\yI$ conforms to $\yS$, based on the desirable and undesirable behaviors defined by regular expressions $D$ e $F$, respectively, according to Proposition~\ref{prop:verifConf}. 
	
	The fault model is represented by automaton $\yA_{T}$ constructed  at line $2$, based on the specification $\yS$ and the languages $(D,F)$. 
	At line $3$ the automaton $\yA_{I}$ is obtained induced by $\yI$. 
	At lines $4$ and $5$, the automaton $\yA_{B}$ is constructed,  which recognizes the common behaviors of the regular languages accepted by the automatons $\yA_{T}$ and $\yA_{I}$. 
	At lines $6$ to $10$, if the set of acceptance states $E$ of the intersection is empty then there is no common behavior between the fault model and the implementation. 
	Therefore, no fault is detected in the IUT and the algorithm declares that $\yI$ conforms to $\yS$ based on $(D,F)$. 
	Otherwise, when at least one fault is detected in $\yI$, the algorithm returns that $\yI$  does not conform to $\yS$ using $(D,F)$.
	\hfill\\
	\begin{breakablealgorithm}
		\label{algoritmo:ConfLing}
		\caption{Verify Language-Based Conformance}
		
		\begin{algorithmic}[1]	
			\Input{LTS $\yS, \yI$ and regex $(D, F)$} 
			\Output{$\yI$  conforms to $\yS$, or not}
			\Function{\textsc{verifyLanguageConformance$(\yS, \yI, D, F)$}}{}
			\State $\yA_{T} = faultModelLanguage(\yS, D, F)$ 
			\State $\yA_{I} = ltsToAutomaton(\yI)$ 
			\State $\yA_{B} =  (S_B,s_0, L, \delta, E)$
			\State $\yA_{B} = intersection(\yA_{T}, \yA_{I})$ 
			
			\If{$E=\emptyset$}
			\return{$\yI$ conforms to $\yS$}
			\Else
			\return{$\yI$ does not conform to $\yS$}
			\EndIf
			\EndFunction					
		\end{algorithmic}
	\end{breakablealgorithm}
	\hfill\\
	
	The language-based fault model is obtained by Algorithm~\ref{algoritmo:TS} based on Lemma~\ref{lema:conjTestComp}. 
	The construction of the automaton that represents the fault model begins with the transformation of the regular expressions $D$ and $F$ into the corresponding automatons that capture their respective languages, at lines $2$ and $3$. 
	The function  $regexToAutomaton$ gets the regular expressions $D$ and $F$ and constructs the automatons $\yA_{D}$ and $\yA_{F}$. 
	Line $4$ constructs the automaton $\yA_{S}$ induced by the LTS $\yS$. 
	At line $5$ the automaton $\yA_{compS}$  is constructed to accept the language complement accepted by $\yA_S$. 
	
	At line $6$ the automaton $\yA_{failD}$ is constructed,  representing the common behaviors between $\yA_{D}$ and $\yA_{compS}$. 
	The intersection contains desirable behaviors that are not present in the specification $\yA_S$. 
	At line $7$ the automaton  $\yA_{failF}$ is constructed representing the common behaviors accepted by $\yA_{F}$ and $\yA_S$. 
	The intersection should include the fault behaviors present in the specification. 
	Finally, at line $8$ the automaton $\yA_{T}$ which corresponds to the union of all fault behaviors is constructed, either by the absence of some desirable behavior that is not in $\yS$ or by the presence of some undesirable behavior that is present in $\yS$. 
	\hfill\\
	\begin{breakablealgorithm}
		\label{algoritmo:TS}
		\caption{Language-Based Fault Model}
		
		\begin{algorithmic}[1]
			\Input{LTS $\yS$ e the regex $(D, F)$} 
			\Output{The fault model automaton}
			
			\Function{\textsc{faultModelLanguage$(\yS, D, F)$}}{}
			\State $\yA_{D} = regexToAutomaton(D)$ 
			\State $\yA_{F} = regexToAutomaton(F)$ 
			\State $\yA_{S} = ltsToAutomaton(\yS)$ 
			\State $\yA_{compS} = complement(\yA_{S})$ 
			\State $\yA_{failD} = intersection(\yA_{D}, \yA_{compS})$
			\State $\yA_{failF} = intersection(\yA_{F}, \yA_{S})$
			\State $\yA_{T} = union(\yA_{failD}, \yA_{failF})$ 
			\return{$\yA_{T}$}	
			\EndFunction
			
		\end{algorithmic}
	\end{breakablealgorithm}
	\hfill\\
	
	\subsection{Everest Tool }\label{sub:interfaceFerramenta}
	
	The conformance verification process using Everest needs  of a standard representation of the models. So LTS/IOLTS models should be modeled in the Aldebaran format \cite{.aut} as a set of transitions.
	\begin{definition}\cite{calame2005}
		Models in Aldebaran format are described by:
		
		des(<initial-state>, <number-of-transitions>, <number-of-states>)
		
		(<ini-state>, <label>, <end-state>)
	\end{definition}
	The initial state (<initial-state>), the number of transitions (<number-of-transitions>) and the number of states (<number-of-states>) are defined in the file header.
	Then, the set of model transitions is defined by the name of the source state (<ini-state>), by the transition label (<label>) and the name of the target state (<end-state>).
	
	Figure~\ref{fig:autModel} presents an example of Aldebaran file ($.aut$)  and Figure~\ref{fig:SubjAut} shows the IOLTS diagram underlying to the Aldebaran model.

	\tikzstyle{mybox} = [draw=black, fill=white!20, very thick,
	rectangle, rounded corners, inner sep=10pt, inner ysep=20pt]
	\begin{figure}[hbt]
		\begin{center}
			\begin{tabular}{@{}cc@{}}
				\subfloat[][IOLTS $\yS$ in Aldebaran format. 
				\label{fig:autModel}]{
					\begin{tikzpicture}
					\node [mybox,inner sep=5pt, inner ysep=5pt] (box){%
						\begin{minipage}{0.18\textwidth}	
						\begin{small}				
						des (s0,9,4)\\
						(s0,?a,s1)\\
						(s0,?b,s3)\\
						(s1,?b,s2)\\
						(s1,!x,s2)\\
						(s1,?a,s3)\\
						(s2,?b,s2)\\
						(s2,!x,s3)\\
						(s3,?b,s0)\\
						(s3,?a,s3)	
						\end{small}				
						\end{minipage}
					};
					\end{tikzpicture}
				}
				&
				\subfloat[IOLTS specification $\yS$.
				\label{fig:SubjAut}]{
					\begin{tikzpicture}[->,>=stealth',shorten >=1pt,auto,node distance=2cm, semithick,initial text =,transform shape]
					
					\node[initial, state]       (s0) {$s_0$};
					\node[state]                (s3) [below of=s0]{$s_3$};
					\node[state]                (s1) [right of=s0] {$s_1$};
					\node[state]               	(s2) [below of=s1]{$s_2$};
					
					\path 
					(s0) edge               node {a}    	(s1)
					edge  [bend left]	node {b}   	(s3)
					(s1) edge               node {a}   	(s3)
					edge               node {b, x}   (s2)
					(s2) edge               node {x}   	(s3)
					edge  [loop below] node {b}   	(s2)   
					(s3) edge  [bend left]  node {b}   	(s0)
					edge  [loop below] node {a}   	(s3);
					\end{tikzpicture}
					
				}
			\end{tabular}
		\end{center}
		\caption{An example of Aldebaran file format. 
			\label{fig:autExemplo} }
	\end{figure}
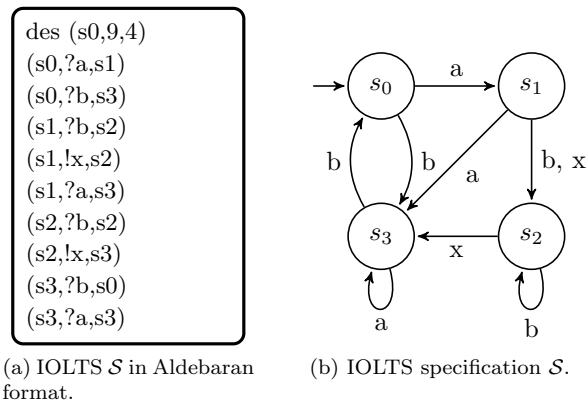
	\vspace{-.8em}
	
\begin{figure}[hbt]
		\centering
		\includegraphics[scale=0.37]{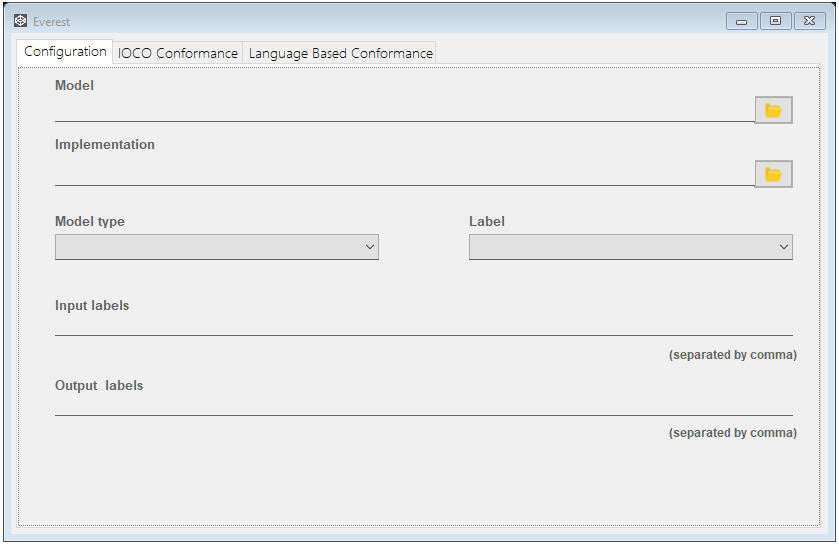}
		\caption{Interface: Configuration.}
		\label{fig:config-everest}
	\end{figure}

	As mentioned, the Everest tool interface consists of three views: configuration; \ioco conformance and language-based conformance.
	Figure~\ref{fig:config-everest} presents the configuration interface, where specification and the implementation models are selected, in the Aldebaran format. 
	When the model type (defined in $Model \; type$) is an LTS, fields $Label$, $Input \; labels$, and $Output \; labels$ are hidden. 
	If IOLTS models are given then we need to inform how the input/output labels are distinguished (field $Label$), informing in the fields $Input \; labels$ and $Output \; labels$,   
	or in the file $.aut$ of the specification/implementation by markers ``!''(output label) e ``?''(input label). 
	
	\vspace{-.8em}
	
\begin{figure}[hbt]
		\centering
		\includegraphics[scale=0.37]{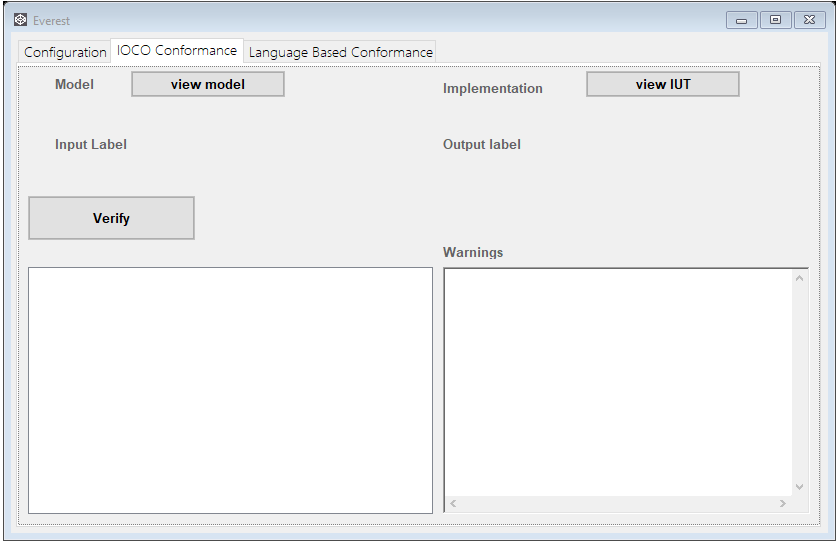}
		\caption{Interface: \ioco verification.}
		\label{fig:ioco-everest}
	\end{figure}

	Figure~\ref{fig:ioco-everest} presents the interface for \ioco conformance verification. 
	In the interface header, the fields fill in the configuration view is kept, as the items $Model$ and $Implementation$ which display the names of the files used as specification and implementation, in addition to the fields $Input \; Label$ and $Output \; Label$, which shows the input and output labels of the models. 
	The buttons $view \; model$ and $view \; IUT$ allow the graphical visualization of the implementation and specification models.
	
	The \ioco conformance verification does not require any additional information,
	therefore, by clicking the button $Verify$, the verdict is textually displayed.
	In case of non-conformance, the tool presents (in the text field below the button), for each test case that detects a fault, a set of paths induced in such a way as to obtain a transition coverage of the specification. 
	The item $Warnings$ informs the incorrectly filled fields in the configuration view, which are essential for the verification to be performed.
	
	\vspace{-.8em}
	
\begin{figure}[hbt]
		\centering
		\includegraphics[scale=0.4]{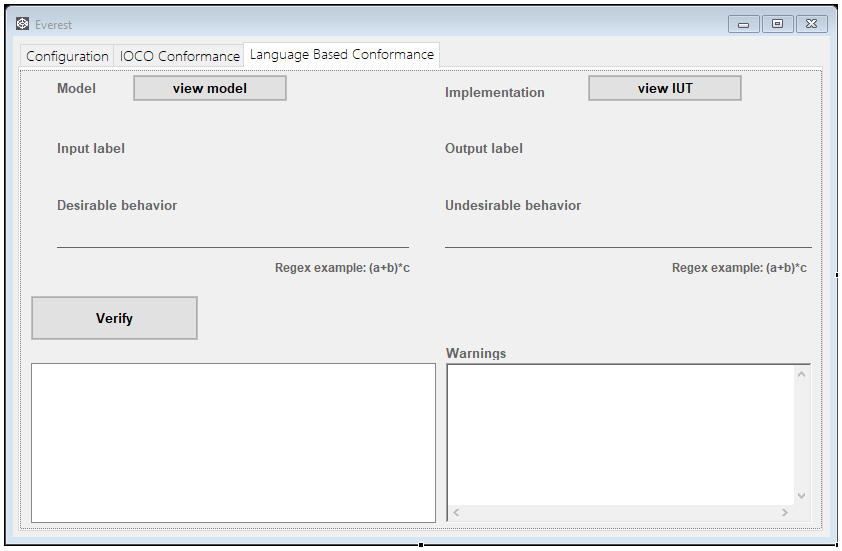}
		\caption{Interface: language-based verification.}
		\label{fig:lang-everest}
	\end{figure}

	Figure~\ref{fig:lang-everest} displays the interface for language-based conformance verification.
	Regular expressions must be informed in the fields $Desirable \; behavior$ and \textit{Undesirable \; behavior}, which specify desirable and undesirable behaviors,  respectively.
	When no regular expression is provided, it is assumed Kleene closure~\cite{sipser2006} over the alphabet, in order to identify faults when models are not isomorphic.
	After, clicking on button $Verify$ the verdict is displayed similarly to the \ioco verification conformance. 
	
	\vspace{-.8em}

	\section{Comparative Analysis of Tools} \label{sub:comparative}
	
	In this section, we perform a comparative analysis between the Everest tool and JTorx. 
	Both tools 
	implement the \ioco theory~\cite{tretmans2008} and 
	JTorx supports both the white-box testing, when the IUT structure is known, and the black-box testing, when the internal structure of the IUT is unknown.
	In the black-box testing, JTorx uses the \textit{Adapter}  module to implement the communication between the IUT and the specification through the TCP or UDP 
	protocol. 
	The conformance verification is accomplished by comparing the outputs expected by the specification and those produced by the implementation.
	
	Everest tool implements the conformance verification on white-box implementations, 
	although the generation of complete test suites for black-box implementations is currently in development.
	In addition to \ioco verification, the Everest tool implements a more general conformance relation that allows specifying fault models and properties through regular languages.
	These languages may express desirable and/or undesirable behaviors over  implementations differently from JTorx that is based only on \ioco conformance. 
	
	In the classical \ioco conformance relation 
	restrictions and properties are imposed on the models, such as underspecified models \cite{belinfante2014} not being allowed, that is, models that are not \textit{input-enabled}. 
	Such restrictions are not imposed on models in the language-based conformance verification.
	Both JTorx and Everest tools support underspecified models, but the states where input labels are not specified, JTorx adds \textit{self-loops}. 
	On the other hand, underspecified models are treated with no change by the Everest tool. 
	
	The Everest test suite generation using the language-based method~\cite{bonifacio2018} is complete over language-based  and \ioco relation for  white-box scenarios.
	The generation algorithms for complete test suites in a black-box testing are under development.
	The JTorx test suite generation is exhaustive with respect to \ioco relation, 
	which implies that it generates very large (infinite) sets of tests, with the same effectiveness on detecting faults using \ioco conformance verification~\cite{goga2001}.

	\section{Practical Application}\label{sub:aplicacao}
	
\begin{figure}[hbt]
	\begin{center}
		\begin{tabular}{@{}cc@{}}			
			\subfloat[][Implementation $\yR$ \label{fig:imp-quase-iso-modificada}]{
				\begin{tikzpicture}[->,>=stealth',shorten >=1pt,auto,node distance=2cm, semithick,initial text =,transform shape]
				
				\node[initial, state]       (s0) {$q_0$};
				\node[state]                (s3) [below of=s0]{$q_3$};
				\node[state]                (s1) [right of=s0] {$q_1$};
				\node[state]               	(s2) [below of=s1]{$q_2$};
				
				\path 
				(s0) edge               node {a}    	(s1)
				edge  [bend left]	node {b}   	(s3)
				(s1) edge               node {a}   	(s3)
				edge               node {b, x}   (s2)
				(s2) edge               node {a}   	(s3)
				edge  [loop below] node {b}   	(s2)   
				(s3) edge  [bend left]  node {b,x}   	(s0)
				edge  [loop below] node {a}   	(s3);
				\end{tikzpicture}
			}\quad
			&
			\subfloat[][Implementation $\yQ$ \label{fig:implementacao_EC_IOCO}]{
				\begin{tikzpicture}[->,>=stealth',shorten >=1pt,auto,node distance=2cm, semithick,initial text =,transform shape]
				
				\node[initial, state]       (q0) {$q_0$};
				\node[state]                (q3) [below of=q0]{$q_3$};
				\node[state]                (q1) [right of=q0] {$q_1$};
				\node[state]               	(q2) [below of=q1]{$q_2$};
				
				\path 
				(q0) edge               node {a}    	(q1)
				edge  [bend left]	node {b}   	(q3)
				(q1) edge               node {a}   	(q3)
				edge               node {b, x}   (q2)
				(q2) edge               node {a,x}   	(q3)
				edge  [loop below] node {b}   	(q2) 
				(q3) edge  [bend left]  node {b}   	(q0)
				edge  [loop below] node {a}   	(q3);
				\end{tikzpicture}
			}	
			
		\end{tabular}
	\end{center}
	\caption{IOLTS Models \label{fig:EC} }
\end{figure}
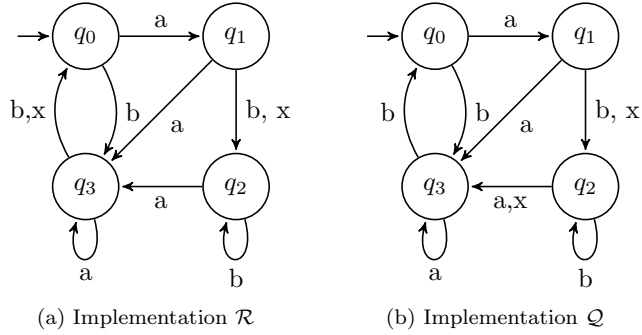

	This section describes some practical testing scenarios applied to the Everest tool and briefly compares with the JTorx tool.
	Let $\yS$  be the IOLTS specification of Figure~\ref{fig:SubjAut} and let  $\yR$ and $\yQ$  be  implementations candidates as depicted in  Figures~\ref{fig:imp-quase-iso-modificada} and~\ref{fig:implementacao_EC_IOCO}. 
	Also let $L_I=\{a,b\}$ and $L_U=\{x\}$ be the input and output alphabets, respectively. 
	All models here are deterministic~\cite{hopcroft2006,utting2007} 
	but we remark that our tool also deals with nondeterministic models.
	
	In the first scenario, we check if IUT $\yR$ conforms to specification $\yS$.  
	Everest tool has returned a non-conformance verdict using  \ioco relation and generated 
	the test suite 
	$\{b,aa,ba,aaa,ab,ax, abb, \\axb\}$. 
	The subset of test cases 
	$\{b,aa,ba,aaa\}$ induces state paths from $s_0$ to  $s_3$ in $\yS$ and from $q_0$ to $q_3$ in $\yR$, where the output $x$ is produced by $\yR$ but $\yS$ does not. 
	Note that $s_3$ in $\yS$ is a quiescent state whence no output is defined on it. 
	The subset $\{ab,ax,abb,axb\}$ induces state paths to state $s_2$ in $\yS$ and $q_2$ in $\yR$. 
	In this case, the output $\delta$  (quiescent label) is produced by IUT $\yR$ whereas $\yS$ produces $x$. 
	That is, a fault is detected according to \ioco relation. 
	Note that both tools modify the formal models by adding \textit{self-loops} labeled with $\delta$~\cite{tretmans1996} on quiescent states. 
	
	The same scenario has been also applied to JTorx tool, resulting in the same verdict, as expected, but it generates the test suite $\{b,ax,ab\}$. 
	Notice that the test suite generated  by JTorx is a subset of the test suite generated by Everest. 
	That is, Everest shows all test cases and associated state paths related to each fault according to a transition cover criteria over the specification, whereas  JTorx returns only one test case per fault. 
	All this information may be useful and aid the tester in the fault mitigation process. 
	
	In the second scenario, checking the IUT $\yQ$ against the specification $\yS$, the language-based conformance verification was able to detect a fault that was not detected 
	by the \ioco conformance relation. 
	We have obtained the fault model using the regular expressions $D=(a|b)^*ax$ and  $F=\emptyset$. 
	Language $D$ clearly expresses behaviors that finish with a stimulus $a$ followed by an output $x$ produced in response. 
	Since the only complete test suite is given by $[(D\cap \overline{otr}(\yS)) \cup (F \cap otr(\yS))]$ and $F \cap otr(\yS) = \emptyset$, so we  check the condition $D\cap \overline{otr}(\yS) \neq \emptyset$, i.e., a fault is detected when behaviors of $D$ are  not present in $\yS$. 
	Everest then results in a verdict of non-conformance and produces the test suite $\{ababax,abaabax\}$ reaching a fault that is not detected by JTorx using the \ioco relation. Following, the details of how the verdicts of \ioco and language-based conformance were obtained from the second scenario according to Everest tool are shown.
	
\begin{figure}[hbt]
		\begin{center}
			\begin{tikzpicture}[->,>=stealth',shorten >=1pt,auto,node distance=1.5cm, semithick,initial text =,transform shape]
			
			\node[initial, state]       (s0) {$s_0$};
			\node[state]                (s3) [below of=s0]{$s_3$};
			\node[state]                (s1) [right of=s0] {$s_1$};
			\node[state]               	(s2) [below of=s1]{$s_2$};
			\node[state,accepting]               	(comp) [punkt, inner sep=3pt,right=1.5cm and 1.5cm  of s2]{$c$};

			\path 
			(s0) edge               node {a}    	(s1)
			edge  [bend left]	node {b}   	(s3)
			edge  [bend left=55]	node {x}   	(comp)
			(s1) edge               node {a}   	(s3)
			edge               node {b, x}   (s2)
			(s2) edge               node {x}   	(s3)
			edge  [loop below] node {b}   	(s2) 
			edge               node {a}  	(comp)
			(s3) edge  [bend left]  node {b}   	(s0)
			edge  [bend right=60]  node[pos=0.5, below] {x}   	(comp)
			edge  [loop left] node [above]{a}   	(s3)
			(comp) edge  [loop right] node [above,pos=0.2]{a, b, x} (comp);
			\end{tikzpicture}			
		\end{center}
		
		\caption{Automaton $\overline{\yA_{S}}$: complement of specification $\yA_{S}$ \label{fig:complement}}
	\end{figure}
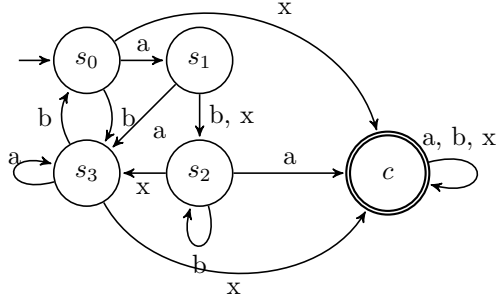

	The specification $\yS$ (Figure~\ref{fig:SubjAut}) and the candidate implementation $\yQ$  (Figure~\ref{fig:implementacao_EC_IOCO}) are IOLTS models which, after being converted into underlying automatons respectively, $\yA_{S}$ and $\yA_{Q}$, have all their states defined as final states. Figure~\ref{fig:complement} displays the complement automaton of the specification. 
	
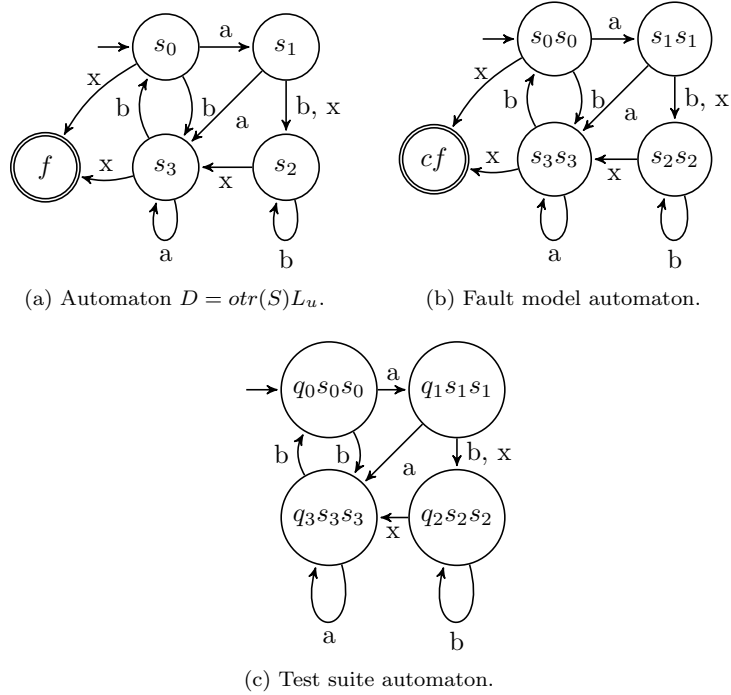
\begin{figure}[hbt]
		\begin{center}
			\begin{tabular}{@{}cc@{}}		
				\subfloat[][Automaton $D = otr(S)L_u$. \label{fig:automatoD_EC_IOCO}]{
					\begin{tikzpicture}[->,>=stealth',shorten >=1pt,auto,node distance=1.6cm, semithick,initial text =]
					
					\node[initial, state]       (s0) {$s_0$};
					\node[state]                (s3) [below of=s0]{$s_3$};
					\node[state]                (s1) [right of=s0] {$s_1$};
					\node[state]               	(s2) [below of=s1]{$s_2$};
					\node[state, accepting]     (falha) [left of= s3]{$f$};

					\path 
					(s0) edge               node {a}    	(s1)
					edge  [bend left]	node {b}   	(s3)
					edge  [bend right=15]	node[above] {x}   (falha)
					(s1) edge               node {a}   	(s3)
					edge               node {b, x}   (s2)
					(s2) edge               node {x}   	(s3)
					edge  [loop below] node {b}   	(s2) 
					(s3) edge  [bend left]  node {b}   	(s0)
					edge  [loop below] node {a}   	(s3)
					edge  [bend left=15]	node[above] {x}   (falha);
					\end{tikzpicture}
				}\quad
				&
				\subfloat[Fault model automaton. \label{fig:modeloFalhaEC_IOCO}]{
					\begin{tikzpicture}[->,>=stealth',shorten >=1pt,auto,node distance=1.6cm, semithick,initial text = ]
					
					\node[initial, state]       (s0) {$s_0s_0$};
					\node[state]                (s3) [below of=s0]{$s_3s_3$};
					\node[state]                (s1) [right of=s0] {$s_1s_1$};
					\node[state]               	(s2) [below of=s1]{$s_2s_2$};
					\node[state, accepting]     (falha) [left of= s3]{$c f$};

					\path 
					(s0) edge               node {a}    	(s1)
					edge  [bend left]	node {b}   	(s3)
					edge  [bend right=15]	node[above] {x}   (falha)
					(s1) edge               node {a}   	(s3)
					edge               node {b, x}   (s2)
					(s2) edge               node {x}   	(s3)
					edge  [loop below] node {b}   	(s2) 
					(s3) edge  [bend left]  node {b}   	(s0)
					edge  [loop below] node {a}   	(s3)
					edge  [bend left=15]	node[above] {x}   (falha);
					\end{tikzpicture}
				}
			\end{tabular}
			
			\subfloat[Test suite automaton. \label{fig:tsEC_IOCO}]{
				\begin{tikzpicture}[->,>=stealth',shorten >=1pt,auto,node distance=1.7cm, semithick,initial text = ]
				
				\node[initial, state]       (q0) {$q_0s_0s_0$};
				\node[state]                (q3) [below of=q0]{$q_3s_3s_3$};
				\node[state]                (q1) [right of=q0] {$q_1s_1s_1$};
				\node[state]               	(q2) [below of=q1]{$q_2s_2s_2$};

				\path 
				(q0) edge               node {a}    	(q1)
				edge  [bend left]	node [left] {b}   	(q3)
				(q1) edge               node {a}   	(q3)
				edge               node {b, x}   (q2)
				(q2)
				edge  [loop below] node {b}   	(q2) 
				edge   node {x}   	(q3)
				(q3) edge  [bend left]  node {b}   	(q0)
				edge  [loop below] node {a}   	(q3);
				\end{tikzpicture}
			}
			
		\end{center}
		\caption{Automatons: \ioco conformance verification \label{fig:EC_IOCO} }
	\end{figure}

	The \ioco conformance verification (Algorithm~\ref{algoritmo:ConfIOCO}) first obtains the underlying automata, $\yA_{S}$ and $\yA_{Q}$, from the IOLTS models.
	The automaton $D$ (Figure~\ref{fig:automatoD_EC_IOCO}) is constructed according to the Algorithm~\ref{algoritmo:ConstroiD}, in order to obtain the fault model of Figure~\ref{fig:modeloFalhaEC_IOCO} through the intersection of the automaton $D$ and the complement of the specification (Figure~\ref{fig:complement}).

	The automaton that represents the test suite (Figure~\ref{fig:tsEC_IOCO}) is obtained by Algorithm~\ref{algoritmo:TS_IOCO} with the intersection between the fault model and the given implementation.
	Since the resulting automaton has no final state,
	the verdict between models is that $\yI$ \ioco conforms to $\yS$.
	
\begin{figure}[hbt]
		\begin{center}
			\begin{tabular}{@{}cc@{}}		
				\subfloat[][Automaton D. \label{fig:automatoD_EC_Ling}]{
					\begin{tikzpicture}[->,>=stealth',shorten >=1pt,auto,node distance=1.8cm, semithick,initial text =,scale=0.8, transform shape]
					
					\node[initial, state]       (d0) {$d_0$};
					\node[state]                (d1) [right of=d0] {$d_1$};
					\node[state, accepting]     (d2) [below of=d0]{$d_2$};

					\path 
					(d0) edge  [loop above] 		node {b}    (d0)
					edge  [bend right = 30]    node {a}   	(d1)
					(d1) edge  [loop above]         node {a}   	(d1)
					edge   				 	node {x}    (d2)
					edge  [bend right = 30]    node {b}    (d0);
					\end{tikzpicture}
					
				}\quad
				&
				\subfloat[Automaton fault model $D \cap \overline{otr}(\yS)$. \label{fig:falhaD_EC_Ling}]{
					\begin{tikzpicture}[->,>=stealth',shorten >=1pt,auto,node distance=1.8cm, semithick,initial text =, scale=0.8, transform shape]
					
					\node[initial, state]       (s0d0) {$s_0d_0$};
					\node[state]                (s1d1) [below of=s0d0]{$s_1d_1$};
					\node[state]                (s3d0) [right of=s0d0] {$s_3d_0$};
					\node[state]               	(s3d1) [right of=s3d0]{$s_3d_1$};
					\node[state]                (s2d0) [right of=s1d1] {$s_2d_0$};
					\node[state]                (s2d2) [below of=s1d1] {$s_2d_2$};
					\node[state,accepting]                (cd2) [right of=s2d0] {$cd_2$};
					\node[state]                (cd1) [below of=cd2] {$cd_1$};
					\node[state]      (cd0) [left of=cd1] {$cd_0$};
					
					\path 
					(s0d0) edge               node {a}   (s1d1)
					edge               node [below] {b}   (s3d0)
					(s1d1) edge               node {a}   (s3d1)
					edge               node {b}   (s2d0)
					edge      node {x}   (s2d2)
					(s3d0) edge               node {a}   (s3d1)		   
					edge   [bend right = 40]  node [above]{b}   (s0d0)	
					
					(s3d1) edge    [loop above] node {a}  (s3d1)
					edge   [bend right = 60]  node [above] {b}   (s0d0)
					edge               node {x}   (cd2)
					(s2d0) edge  [loop right] node [above] {b} (s2d0)
					edge    node {a}   (cd1)
					
					(cd1) edge   [loop right]  node {a}   	(cd1)
					edge [bend right = 15]   node [above] {b}  	(cd0)					
					edge  node {x}   	(cd2)
					
					(cd0) edge  [bend right = 15] node [below] {a} (cd1)
					edge  [loop above] node [left] {b}(cd0);
					
					\end{tikzpicture}
				}
			\end{tabular}
			
			\begin{tabular}{@{}c@{}}
				\subfloat[][Automaton test suite.  \label{fig:tsEC_Ling}]{
					\begin{tikzpicture}[->,>=stealth',shorten >=1pt,auto,node distance=2cm, semithick, initial text =, scale=0.8, transform shape]
					
					\node[initial, state]       (s0d0q0) {$s_0d_0q_0$};
					\node[state]                (s1d1q1) [right of=s0d0q0]{$s_1d_1q_1$};
					\node[state]                (s2d2q2) [right of=s1d1q1]{$s_2d_2q_2$};
					\node[state]                (s3d0q3) [below of=s0d0q0]{$s_3d_0q_3$};
					\node[state]                (s3d1q3) [right of=s3d0q3]{$s_3d_1q_3$};							
					\node[state]                (cd1q3) [right of=s3d1q3]{$cd_1q_3$};
					\node[state]                (s2d0q2) [right of=cd1q3]{$s_2d_0q_2$};
					\node[state]                (cd0q2) [below of=s3d0q3]{$cd_0q_2$};
					\node[state]                (cd1q1) [right of=cd0q2]{$cd_1q_1$};			
					\node[state]                (cd0q0) [right of=cd1q1]{$cd_0q_0$};					
					\node[state]                (cd0q3) [right of=cd0q0]{$cd_0q_3$};		
					\node[state, accepting]     (cd2q2) [below of=cd1q1]{$cd_2q_2$};
					
					\path 
					(s0d0q0) edge               node {a}   (s1d1q1)									
					edge           [bend right = 15]    node [left]{b}   (s3d0q3)	
					(s1d1q1) edge               node {a}   (s3d1q3)	
					edge   [bend left = 5]           node {b}   (s2d0q2)	
					edge               node {x}   (s2d2q2)
					(s3d0q3) edge    [bend right = 15]           node [right]{b}   (s0d0q0)	
					edge               node {a}   (s3d1q3)
					(s3d1q3) edge        [loop right]       node [above]{a}   (s3d1q3)	
					edge               node {b}   (s0d0q0)		
					(s2d0q2) edge               node {a}   (cd1q3)		
					edge     [loop above]            node [right]{b}   (s2d0q2)	
					(cd1q3) edge    [loop above]            node [left]   {a} (cd1q3)
					edge   node {b}   (cd0q0)	
					(cd0q0) edge   node {a}   (cd1q1)		
					edge  [bend right = 15] node [below] {b}   (cd0q3)		
					(cd0q3) edge   node {a}   (cd1q3)
					edge [bend right = 15]  node [above]{b}   (cd0q0)
					(cd1q1) edge   node {a}   (cd1q3)
					edge  node {x}   (cd2q2)
					edge   node {b}   (cd0q2)
					(cd0q2) edge   node {a}   (cd1q3)
					edge [loop above]            node {b}   (cd0q2);

					\end{tikzpicture}
				}
			\end{tabular}
		\end{center}
		\caption{Automaton: language-based conformance verification. \label{fig:EC_Ling} }
	\end{figure}
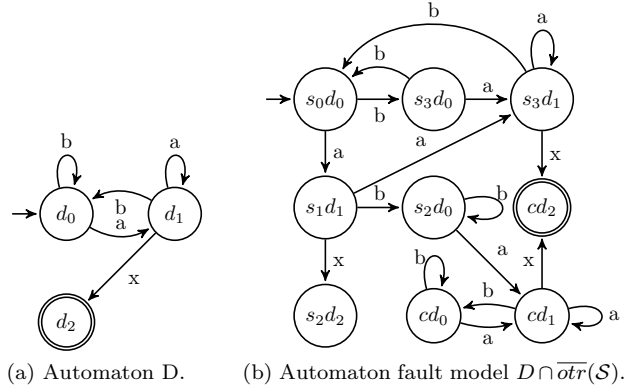
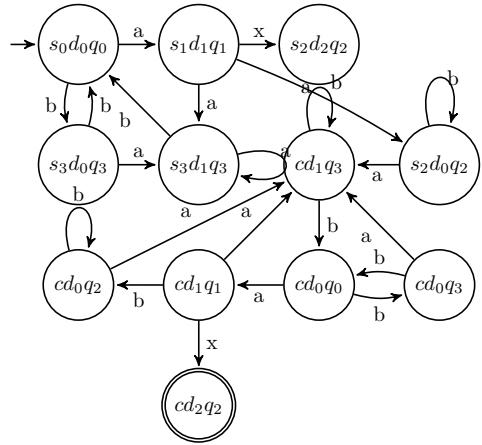

	In language-based conformance verification, as well as
	in the \ioco verification,
	the underlying automatons $\yA_{S}$ and $\yA_{Q}$, are obtained from the IOLTS models,
	according to Figure~\ref{fig:EC} and Figure~\ref{fig:autExemplo}. 
	From regular expression $(a|b)^*ax$ the automaton (Figure~\ref{fig:automatoD_EC_Ling}) which accepts the respective language is also obtained. 
	Since the fault model is given by $[D \cap \overline{otr}(\yS)]\cup[(F \cap otr(\yS))]$ (Algorithm~\ref{algoritmo:TS}) and 
	no undesirable behavior $F$ is informed, therefore
	$F \cap otr(\yS) = \emptyset$, so fault behaviors are reduced to
	$D \cap \overline{otr}(\yS)$. 
	The automaton that represents the fault model is
	illustrated in
	Figure~\ref{fig:falhaD_EC_Ling}.

	Finally, the automaton that represents the test suite is illustrated in Figure~\ref{fig:tsEC_Ling} and obtained by Algorithm~\ref{algoritmo:ConfLing}. 
	Note that this automaton contains a final state,
	indicating that the words accepted by the
	automaton are part of the test suite that reveals the faults and, consequently, the non-conformity between the models.
	The test suite generated by the Everest tool are $\{ababax,abaabax\}$.
	
\section{Conclusion}\label{sub:conclusao}

Testing of reactive systems is an important and complex activity in the development process for systems of this nature. 
The complexity of such systems and, consequently,  the complexity of the testing task  require high costs and resources in software development. 
Therefore, automation of the testing activity has become  essential 
and several studies have addressed the testing of reactive systems~\cite{bonifacio2018,simao2014,tretmans2008,aichernig2015,anand2013}. 
More precisely, many works have focused on the conformance checking between IUTs and specifications 
to guarantee more reliability. 

In this work, we have developed an automatic tool for checking conformance on asynchronous reactive systems. 
We have implemented not only the  more general relation based on regular languages but also the classical \ioco theory.
We could observe that the Everest conformance verification process based on regular languages can find faults that JTorx could not 
detect using the classical \ioco relation.
In the practical applications, we could see that JTorx yielded a verdict of conformance whereas Everest could find a fault for the same scenario.

In spite of the studies over these distinct approaches of conformance  theory, 
we remark that the main contribution of this work is the tool development together with the designed algorithms, providing an intuitive graphical interface  either for experts in the research area or for beginners with no specific  knowledge over the conformance theories. 

A new module of Everest tool is already being developed to provide the test suite generation in a black-box setting.  
We also intend to perform more experiments and comparative studies with similar tools from the literature in order to give a more precise analysis related to the conformance checking issue, specially, with regard the usability and performance of these tools.

\end{document}